\documentclass{article}

\usepackage[a4paper, margin=3.3cm]{geometry}

\usepackage{amsmath, amsfonts, amssymb, amsthm}
\usepackage{xspace}
\usepackage{subfig}
\usepackage{upgreek}
\usepackage{wrapfig}

\renewcommand{\geq}{\geqslant}

\renewcommand{\epsilon}{\varepsilon}
\renewcommand{\Delta}{\Updelta}

\newcommand{\TRUE}{\textsc{True}\xspace}
\newcommand{\FALSE}{\textsc{False}\xspace}

\newcommand{\indexing}{\textsc{Indexing}\xspace}
\newcommand{\equality}{\textsc{Equality}\xspace}
\newcommand{\disjointness}{\textsc{Disjointness}\xspace}

\newcommand{\SigmaP}{\ensuremath{\Sigma_{\textup{P}}}}
\newcommand{\SigmaT}{\ensuremath{\Sigma_{\textup{T}}}}
\newcommand{\local}{\ensuremath{\textsc{LocalPM}}}

\newcommand{\wildcard}{\ensuremath{\star}}

\theoremstyle{plain}
\newtheorem{theorem}{Theorem}[]
\newtheorem{lemma}[theorem]{Lemma}
\newtheorem{corollary}[theorem]{Corollary}

\newtheorem{example}[theorem]{Example}

\theoremstyle{definition}
\newtheorem{definition}[theorem]{Definition}

\newtheorem*{proviso*}{Proviso}

\title{Space Lower Bounds for Online Pattern Matching}
\author{
    Rapha\"{e}l Clifford,%
    \thanks{~University of Bristol, Dept.\@ of Computer Science, Bristol, UK}
    \and Markus Jalsenius,%
    \footnotemark[1]\\
    \and Ely Porat,%
    \thanks{~Bar-Ilan University, Dept.\@ of Computer Science, Ramat-Gan, Israel}
    \and Benjamin Sach%
    \footnotemark[1]}

\date{}

\begin{document}

\maketitle

\begin{abstract}
    We present space lower bounds for online pattern matching under a number of different distance measures. Given a pattern of length $m$ and a text that arrives one character at a time, the online pattern matching problem is to report the distance between the pattern and a sliding window of the text as soon as the new character arrives. We require that the correct answer is given at each position with constant probability. We give $\Omega(m)$ bit space lower bounds for $L_1$, $L_2$, $L_{\infty}$, Hamming, edit and swap distances as well as for any algorithm that computes the cross-correlation/convolution. We then show a dichotomy between distance functions that have wildcard-like properties and those that do not. In the former case which includes, as an example, pattern matching with character classes, we give $\Omega(m)$ bit space lower bounds. For other distance functions, we show that there exist space bounds of $\Omega(\log m)$ and $O(\log^2 m)$ bits. Finally we discuss space lower bounds for non-binary inputs and show how in some cases they can be improved.
\end{abstract}

\section{Introduction}
We combine existing results with new observations to present an overview of space lower bounds for online pattern matching.  Given a pattern that is provided in advance and a text that arrives one character at a time, the online pattern matching problem is to report the distance between the pattern and a sliding window of the text as soon as the new character arrives. In this formulation, the pattern is processed before the first text character arrives and once processed, the pattern is no longer available to the algorithm unless a copy is explicitly made.

This problem has recently gained a great deal of interest with breakthrough results
 given for exact matching and pattern matching under bounded
Hamming distance ($k$-mismatch)~\cite{Porat:09}.
For both problems it was shown that space sublinear in the size of the pattern is sufficient to give the correct answer at every alignment with high probability.  These remarkable results immediately raise a number of significant unresolved questions. The first is for which other distance measures between strings might sublinear space randomised online algorithms be achievable and it is this question which we address here.

Our presentation is divided between what we term local and non-local online pattern matching problems. In the former case the distance function between a pattern $P$ of length $m$ and an $m$-length substring of the text $T$, starting at position $i$, is defined by
\begin{equation*}
    \local_{(\oplus,\Delta)}(P,T) = \bigoplus_{j=0}^{m-1}\Delta(P[j],T[i+j])\,,
\end{equation*}
where $\oplus$ and $\Delta$ are both binary operators. In Section~\ref{sec:addition} we show $\Omega(m)$ bit space lower bounds for online pattern matching for the local problems of $L_{1}$, $L_{2}$, and Hamming distance as well as for any algorithm that computes the cross-correlation/convolution.

We then go on to show in Section~\ref{sec:wildcard} a  space dichotomy for local online pattern matching problems of the form $d(i)= \bigwedge_{j=0}^{m-1}\Delta(P[j],T[i+j])$
%
%
where the range of $\Delta$ is $\{\TRUE,\FALSE\}$. Where the distance function $\Delta$ has wildcard-like properties (qv.~Section~\ref{sec:wildcard}), we give an $\Omega(m)$ space lower bound. Where it does not, we have $\Omega(\log{m})$ and $O(\log^2{m})$ space bounds.  This implies, for example, that online pattern matching with character classes~\cite{LR:2009} requires linear space.

In Section~\ref{sec:otherbool} we go on to consider all eight possible binary Boolean associative operators and give a complete classification in terms of their known upper and lower space bounds.  One consequence is that determining if there is an exact ``non-match'', where the Hamming distance is the same as the pattern length, requires linear space in our online model. This bound also holds if, for example, only the parity of the Hamming distance is required.  In Section~\ref{sec:infty} we then show how our techniques can be used to give linear space lower bounds for  $L_{\infty}$ online pattern matching.  In Section~\ref{sec:nonbinary} we discuss a possible approach to space lower bounds for inputs with large alphabets, focussing on the Hamming distance problem. Finally, in Section~\ref{sec:non-local} we explore non-local problems and show $\Omega(m)$ bit space lower bounds for both online edit and swap distance.

\section{Preliminaries and related work}

Let $\SigmaP$ and $\SigmaT$ denote the pattern and text alphabet, respectively. We say that $\local_{(\oplus,\Delta)}$ is \emph{text independent} with respect to the pattern $P$ if the value of $\local_{(\oplus,\Delta)}$ is a constant independent of $T$. We say that $\local_{(\oplus,\Delta)}$ is \emph{pattern independent} with respect to a pattern $P$ if there is a function $\Delta'$ such that $\Delta(x,y)=\Delta'(y)$ for all $(x,y)\in P\times\SigmaT$.

\begin{example}
    Let $\SigmaP=\left\{ x,y,z\right\} $, $\SigmaT=\left\{ a,b,c\right\} $, $\oplus$ be the Boolean AND-operator and $\Delta$ be defined according to the table in Figure~\ref{fig:example-invalid}, where 1 is \TRUE and 0 is \FALSE. We can see that $\local_{(\wedge,\Delta)}$ is text independent with respect to the pattern $P=xxyyxzxx$ as it always outputs $0$. It is also pattern independent with respect to $P=yyzyyzzy$ as $\Delta(y,\alpha) = \Delta(z,\alpha)$ for all $\alpha\in\SigmaT$. In fact, for this particular definition of $\Delta$, $\local_{(\wedge,\Delta)}$ is either text or pattern independent with respect to any pattern $P$.
    \begin{figure}[t]
        \begin{equation*}
            \begin{array}{c|ccc}
                \Delta & a & b & c\\
                \hline x & 0 & 0 & 0\\
                y & 0 & 1 & 1\\
                z & 0 & 1 & 1
            \end{array}\,
        \end{equation*}
        \caption{\label{fig:example-invalid}An example of $\Delta$ such that $\local_{(\wedge,\Delta)}$ is invalid (either text or pattern independent with respect to any pattern $P$).}
    \end{figure}
\end{example}

Suppose that $\local_{(\oplus,\Delta)}$ is text independent with respect to a pattern $P$. Then any algorithm for $\local_{(\oplus,\Delta)}$ on $P$ requires at most $O(1)$ space after preprocessing $P$. If $\local_{(\oplus,\Delta)}$ is pattern independent with respect to $P$ then $\local_{(\oplus,\Delta)}$ does not depend on the pattern and is outside the scope of this paper.

We say that $\local_{(\oplus,\Delta)}$ is \emph{invalid} if, for every pattern $P$, it is either text or pattern independent with respect to $P$. $\local_{(\oplus,\Delta)}$ is \emph{valid} if it is not invalid. The problem $\local_{(\wedge,\Delta)}$ in the previous example is therefore invalid. We will only consider from this point pattern matching problems $\local_{(\oplus,\Delta)}$ which are valid, and ignore patterns for which $\local_{(\oplus,\Delta)}$ is pattern or text independent.

Our focus is on online pattern matching algorithms which output correct answers with constant probability.  We are not aware of previous work that considers randomised lower bounds for this specific type of problem. There is however now a considerable literature on communication complexity and on streaming algorithms for single input streams, including those that process a sliding window of the input (see e.g.~\cite{DGIM:02})). This previous streaming work has typically focussed on deterministic or randomised bounds for finding approximate rather than exact solutions.  Quantum lower and classical upper bounds for the communication complexity of Hamming distance in more general models than we consider were given previously~\cite{HSZZ:06}. A linear lower bound for the randomised communication complexity of the inner product of two binary vectors is given in~\cite{CG:1985}.   The dichotomy presented in Section~\ref{sec:wildcard} and in particular the concept of  a matching relation that includes wildcard matching, although in a different setting and with different terminology, is similar to a time complexity dichotomy given previously by Muthukrishnan and Ramesh~\cite{MR:95}. On the topic of swap matching in Section~\ref{sec:non-local}, we note that in~\cite{AALLL:00}, the existence of a reduction for time rather than space, from Boolean convolutions to string matching with swaps is claimed without proof.

\section{Communication complexity problems}

Our results are based on reductions from various one-way randomised communication complexity problems with known lower bounds. We list the relevant problems below. In a one-way randomised communication model, only Alice can send messages to Bob and Bob must output the correct answer with probability at least $2/3$. Note that the value $2/3$ is inconsequential: any probability strictly greater than $1/2$ can be amplified to a constant arbitrary close to 1. We assume private randomness.

\begin{definition}
    The \equality problem in one-way communication complexity is defined as follows. Alice has a string $X\in\{0,1\}^{m}$ and Bob has a string $Y\in\left\{ 0,1\right\} ^{m}$. Bob must determine whether $X=Y$. The communication complexity is $\Theta(\log m)$ bits~\cite{Yao:79}.
\end{definition}

\begin{definition}
    \label{def:indexing}
    The \indexing problem in one-way communication complexity is defined as follows. Alice has a string $X\in\{0,1\}^{m}$ and Bob has an index $n\in\{0,\ldots m-1\}$. Bob must find $X[n]$. The problem is known to have an $\Omega(m)$ bit lower bound (see~\cite{JKS:08} for an elementary proof).
\end{definition}

\section{Addition}\label{sec:addition}

In this section we consider the problem $\local_{(+,\Delta)}$, where $+$ is standard addition and the range of $\Delta$ is a subset of the integers. That is, the distance function is
\begin{equation*}
    d(i)=\sum_{j=0}^{m-1}\Delta(P[j],T[i+j])\,.
\end{equation*}

\begin{theorem}
    $\local_{(+,\Delta)}$ requires $\Omega(m)$ bits of space.
\end{theorem}
\begin{proof}
    Since $\local_{(+,\Delta)}$ is not text independent, there must exist characters $x\in\SigmaP$ and $a,b\in\SigmaT$ such that $\Delta(x,a)\neq\Delta(x,b)$. We reduce from \indexing: Alice has a string $T=\{a,b\}^m$ and Bob has an index $n$. Alice initialises a pattern matching algorithm $A$ on the pattern $P=\{x\}^m$ and feeds in her string $T$. Then she sends the internal state of $A$ to Bob, who feeds in $n$ copies of the symbol $a$. Let $d$ be the output after those $a$s. Bob then feeds in another $a$. Let $d'$ be the output. If $d=d'$ then $A[n]=a$. If $d\neq d'$ then $A[n]=b$. If the probability of error per output is bounded by a constant $c<1/4$, then the union bound for error on two outputs is $2c$, giving the \indexing problem an error probability of at most $2c<1/2$.
\end{proof}

\begin{corollary}
    \label{cor:standard-addition}
    Computing the $L_{1}$, $L_{2}$ and Hamming distances, as well as the convolution, require $\Omega(m)$ bits of space.
\end{corollary}

\section{Conjunction} \label{sec:wildcard}

In this section we consider $\local_{(\wedge,\Delta)}$, where $\wedge$ is the Boolean AND-operator and the range of $\Delta$ is $\{0,1\}$ (where 0 denotes \FALSE and 1 denotes \TRUE). There are several natural pattern matching problems that fall under this category, for example, exact matching, matching with wildcards and exact matching with character classes.

The function $\Delta$ can be represented with a $0/1$-matrix $M_{\Delta}$, where the rows and columns correspond to the symbols in $\SigmaP$ and $\SigmaT$, respectively. Thus, the entry $(i,j)=\Delta(i,j)$. The $2$$\times$$2$ matrix in Figure~\ref{fig:bad-matrix} will play an important role, and we call it the \emph{wildcard matrix}.

\begin{figure}[t]
    ~
    \hfill
    \begin{minipage}[b]{0.5\linewidth}
        \begin{equation*}
            \left[
            \begin{array}{cc}
                1 & 1\\
                1 & 0
            \end{array}
            \right]
            \qquad \qquad
            \left[
            \begin{array}{cc}
                0 & 0\\
                0 & 1
            \end{array}
            \right]
        \end{equation*}
        \vspace{1pt}
        \caption{\label{fig:bad-matrix}The wildcard matrix (left) and negated wildcard matrix (right).}
    \end{minipage}
    ~
    \hfill
    ~
    \begin{minipage}[b]{0.3\linewidth}
         \begin{equation*}
             \begin{array}{c|cc}
                 \Delta & a & b\\
                 \hline \wildcard & 1 & 1\\
                 x & 1 & 0
             \end{array}\,.
         \end{equation*}

         \caption{\label{fig:and-linear}$\Delta$ in the proof of Theorem~\ref{thm:and-linear}.\\~}
          \vspace{-18.25pt}
    \end{minipage}
    \hfill
    ~
\end{figure}

We say that $M_{\Delta}$ contains the wildcard matrix if it is a submatrix of $M_{\Delta}$ under some permutation of the rows and columns.

We demonstrate the following dichotomy for $\local_{(\wedge,\Delta)}$. If $M_{\Delta}$ contains the wildcard matrix, then $\local_{(\wedge,\Delta)}$ is solvable in $\tilde{\Theta}(m)$ bits of space, otherwise it is solvable in $\tilde{\Theta}(1)$ bits of space. The first class is equivalent to pattern matching with wildcards, and the second class is equivalent to exact matching. Note that both dichotomies are decidable due to the simple characteristic of the function $\Delta$.

\begin{theorem}
    \label{thm:and-linear}
    If $M_{\Delta}$ contains the wildcard matrix, then $\local_{(\wedge,\Delta)}$ requires $\Omega(m)$ bits of space.
\end{theorem}
\begin{proof}
    Suppose that $\wildcard,x\in\SigmaP$ ($\wildcard$ represents a wildcard symbol) and $a,b\in\SigmaT$ such that $\Delta$ is specified according to Figure~\ref{fig:and-linear}. We reduce from the \indexing problem, in which Alice has an $m$-length bit string $X\in\{\wildcard,x\}^{m}$ and Bob has an index $n\in\{0,\ldots m-1\}$. Let the pattern $P$ be the string $X$. Let $A$ be any algorithm that solves $\local_{(\wedge,\Delta)}$ on the pattern $P$. Alice sends the internal state of $A$ to Bob, who feeds the algorithm with the $m$-length string that has the symbol $a$ at every position except for at position $n$ where the symbol is $b$. The output is \TRUE iff $X[n]=\wildcard$.
\end{proof}

\noindent The following lemma will be useful for the next two theorems (see Figure~\ref{fig:identity}).

\begin{lemma}
    \label{lem:identity}
    Let $M'_{\Delta}$ be the matrix obtained from $M_{\Delta}$ by first removing copies of identical rows and columns, keeping only rows and columns that are distinct in $M_{\Delta}$, and then removing any row or column that contains only zeros. If $M_{\Delta}$ does not contain the wildcard matrix, then $M'_{\Delta}$ is the identity matrix, under some permutation of rows and columns.
\end{lemma}
\begin{proof}
    Suppose that $M_{\Delta}$ does not contain the wildcard matrix. Let $M'_{\Delta}$ be obtained from $M_{\Delta}$ according to the statement of the lemma. We will show that every column and every row of $M'_{\Delta}$ contains exactly one~1.

    First we show that every row of $M'_{\Delta}$ must contain at least one~1. Suppose that some row $r$ of $M'_{\Delta}$ contains only 0s. Since zero-rows of $M_{\Delta}$ were removed and one copy of each column remains after the removal process, it is not possible that all columns in which row $r$ is 1 were removed. We now show that $M'_{\Delta}$ cannot contain a row $r$ with two or more 1s. Without loss of generality, assume that there is a 1 in columns $i$ and $j$ of row $r$. Since $M_{\Delta}$ does not contain a wildcard matrix, the elements of columns $i$ and $j$ must both be either 0 or 1 in every row. Thus, columns $i$ and $j$ are identical, and one of them must have been removed, contradicting the fact that there are two 1s in row $r$ of $M'_{\Delta}$. In order to show that every column of $M'_{\Delta}$ contains exactly one 1, we use the exact same argument as for the rows. Thus, $M'_{\Delta}$ is the identity matrix, under some permutation of rows and columns. (See Figure~\ref{fig:identity} for an illustration of the lemma).
\end{proof}

\begin{figure}[t]
       \vspace{-5pt}
       \begin{equation*}
            \begin{array}{c|cccccc}
                M_{\Delta} & a & b & c & d & e & f\\
                \hline v & 0 & 1 & 0 & 1 & 1 & 0\\
                w & 0 & 0 & 0 & 0 & 0 & 0\\
                x & 1 & 0 & 1 & 0 & 0 & 0\\
                y & 0 & 1 & 0 & 1 & 1 & 0\\
                z & 1 & 0 & 1 & 0 & 0 & 0
            \end{array}
            \qquad
            \begin{array}{c|cccccc}
                M'_{\Delta} & a & b & c & d & e & f\\
                \hline v & 0 & 1 & - & - & - & -\\
                w & - & - & - & - & - & -\\
                x & 1 & 0 & - & - & - & -\\
                y & - & - & - & - & - & -\\
                z & - & - & - & - & - & -
            \end{array}
            \qquad
            \begin{array}{c|cc}
                \textup{{Id.}} & a & b\\
                \hline x & 1 & 0\\
                v & 0 & 1\\
                \\
                \\
                \\
            \end{array}
        \end{equation*}
        \caption{\label{fig:identity}An illustration of Lemma~\ref{lem:identity}.\\~}
\end{figure}

\begin{theorem}
    \label{thm:and-log}
    If $M_{\Delta}$ does not contain the wildcard matrix, then $\local_{(\wedge,\Delta)}$ requires $\Omega(\log m)$ bits of space.
\end{theorem}
\begin{proof}
    We reduce from the \equality problem, where Alice has a string $X\in\left\{ 0,1\right\} ^{m}$ and Bob has a bit string $Y\in\left\{ 0,1\right\} ^{m}$. Since $M_{\Delta}$ doesn't contain the wildcard matrix and as we only consider problems $\local_{(\wedge,\Delta)}$ that are valid, it follows from Lemma~\ref{lem:identity} that there must exist $x,y\in\SigmaP$ and $a,b\in\SigmaT$ such that $\Delta$ is according to Figure~\ref{fig:and-log}.
    Let $P$ be the $m$-length pattern obtained from $X$ by replacing every 0 with $x$ and every 1 with $y$. The $m$-length text $T$ is obtained similarly from $Y$ by replacing every 0 with $a$ and every 1 with $b$. For any algorithm $A$ that solves $\local_{(\wedge,\Delta)}$ on the pattern $P$, Alice sends the internal state of $A$ on pattern $P$ to Bob, who feeds $A$ with $T$. The output is \TRUE iff $X=Y$.
\end{proof}

\begin{theorem}
    \label{lem:and-exact}
    If $M_{\Delta}$ does not contain the wildcard matrix, then $\local_{(\wedge,\Delta)}$ can be solved in $O(\log^{2}m)$ bits of space.
\end{theorem}
\begin{proof}
    We will describe an algorithm for solving $\local_{(\wedge,\Delta)}$ which uses the exact matching algorithm by Porat and Porat~\cite{Porat:09}, which runs in space $O(\log m)$ words, which is $O(\log^{2}m)$ bits of space (under the word-RAM model). In order to use the exact matching algorithm (as a ``black box'') we must ensure that we do not feed it with distinct symbols that are identical under $\Delta$. In other words, we can think of $\Delta$ specifying character classes, and for each class we want to use one representative symbol. We formalise this below.

    We make the very reasonable assumption that the alphabets $\SigmaP$ and $\SigmaT$ are both enumerable and that we can iterate through every symbol of $\Sigma_{P}$ and $\SigmaT$, respectively, in no more than $O(\log m)$ bits of space. Let the order by which we iterate through the alphabets describe an ordering of the symbols in $\SigmaP$ and $\SigmaT$. We say that the symbol $x\in\SigmaP$ is \emph{smaller} than $y\in\SigmaP$ if $x$ appears before $y$ when iterating through $\SigmaP$. We use the same notation for the symbols of $\SigmaT$. We say that two symbols $x,y\in\SigmaP$ are \emph{equivalent} if $\Delta(x,a)=\Delta(y,a)$ for all $a\in\SigmaT$. Similarly, $a,b\in\SigmaT$ are equivalent if $\Delta(x,a)=\Delta(x,b)$ for all $x\in\SigmaP$. We define the \emph{smallest equivalent} symbol of $x\in\SigmaP$ to be the symbol $y\in\SigmaP$ such that $y$ is equivalent to $x$ and no other symbol equivalent to $x$ is smaller than $y$. The notion of smallest equivalent symbol is defined similarly on $\SigmaT$.

    Let $\SigmaP'\subseteq\SigmaP$ be the set of all symbols $x\in\SigmaP$ such that the smallest equivalent symbol of $x$ is $x$ itself. We do not include any symbol $x$ in $\SigmaP'$ such that $\Delta(x,a)=0$ for all $a\in\SigmaT$. Similarly, let $\SigmaT'\subseteq\SigmaT$ be the set of all symbols $a\in\SigmaT$ such that the smallest equivalent symbol of $a$ is $a$ itself. We do not include any symbol $a$ in $\SigmaT'$ such that $\Delta(x,a)=0$ for all $x\in\SigmaP$. By Lemma~\ref{lem:identity} we have that $\Delta$ on $\SigmaP'$ and $\SigmaT'$ is represented by an identity matrix under some permutation of the rows and columns. In the example of Figure~\ref{fig:identity}, $\SigmaP'=\{x,v\}$ and $\SigmaT'=\{a,b\}$. We will ensure that we use the exact matching algorithm of~\cite{Porat:09} only on $\SigmaP'$ and $\SigmaT'$ (i.e., normal exact pattern matching).

    Given a symbol $x\in\SigmaP$, we can find its smallest equivalent symbol by iterating through every symbol $y\in\SigmaP$ and for each $y$, we iterate through all $a\in\SigmaT$ to check whether $\Delta(x,a)=\Delta(y,a)$. Similarly we can find the smallest equivalent symbol of any symbol in $\SigmaT$.

    Let $P$ be the pattern. We may assume that $P$ does not contain a symbol $x$ for which $\Delta(x,a)=0$ for all $a\in\SigmaT$. If it does, the output is always 0. Before we preprocess the pattern, we replace every symbol with its smallest equivalent symbol. Then we preprocess the pattern using the fingerprint technique described in~\cite{Porat:09}. Now we run the exact matching algorithm with the following additional step. When a new symbol $a$ arrives, we replace it with its smallest equivalent symbol. The only caveat we must take care of is the situation when $\Delta(x,a)=0$ for all $x\in\SigmaP$. We can detect this case by iterating through the symbols of $\SigmaP$. As long as $a$ is present in the last $m$ characters of the stream, the output is zero. We use a flag to keep track of this.
\end{proof}

We now show how these results can be applied to a specific pattern matching problem that has not been considered in the online setting before. The pattern matching with character classes problem allows a set of characters to be defined for each position in the pattern~\cite{LR:2009}.  A character in the text matches a set at a pattern position if it is contained within it.  This is a generalisation of exact matching where each set would contain only one character. Using Theorems~\ref{thm:and-linear},~\ref{thm:and-log} and~\ref{lem:and-exact} we can determine precisely when this problem can and cannot be solved online in sublinear space.

\begin{corollary}
    Online pattern matching with character classes requires $\Omega(m)$ bits of space in the worst case. However, where the character classes define a matching relation $\Delta$ which does not contain the wildcard matrix (see the example in Figure~\ref{fig:identity}), $O(\log^2{m})$ bits suffice.
\end{corollary}

\begin{figure}[t]
        \begin{equation*}
            \begin{array}{c|cc}
                \Delta & a & b\\
                \hline x & 1 & 0\\
                y & 0 & 1
            \end{array}
        \end{equation*}
        \caption{\label{fig:and-log}$\Delta$ in the proof of Theorem~\ref{thm:and-log}.}
\end{figure}

\section{Other Boolean operators} \label{sec:otherbool}

In the previous section we demonstrated a dichotomy for $\local_{(\oplus,\Delta)}$, where $\oplus$ is the AND-operator. Here we will complete the classification of Boolean operators. There are eight associative Boolean operators $a\oplus b$:

\medskip
\noindent
\hfill
\textbf{1.} \TRUE \hfill
\textbf{2.} \FALSE \hfill
\textbf{3.} $a$ \hfill
\textbf{4.} $b$ \hfill
\textbf{5.} $a\wedge b$ \hfill
\textbf{6.} $a\vee b$ \hfill
\textbf{7.} $a=b$ \hfill
\textbf{8.} $a\neq b$ \hfill

\medskip
The operators \TRUE and \FALSE are trivial; the output is either always \TRUE or \FALSE. The operator $a\oplus b=b$ is also easy; the output is always $\Delta(P[m-1],t)$, where $t$ is the last received symbol of the text stream.

The operator $a\oplus b=a$ is on the other hand more demanding. Here the output is $\Delta(P[0],t)$, where $t$ is the $m$th last symbol received from the text stream. The pattern matching algorithm must therefore remember $m$ received characters of the stream. More precisely, we see that $\Omega(m)$ bits of space is necessary by reducing from the \indexing problem: Alice first feeds her array (text) into the pattern matching algorithm, for which $P[0]$ is a character that can distinguish between the characters of Alice's array. She then sends the internal state to Bob, who feeds in $n$ symbols in order to determine the value at index $n$ of Alice's array.

The OR-operator $\vee$ is equivalent to $\wedge$ under De Morgan's laws: negate the outputs from $\Delta$ and negate the output from the pattern matching algorithm. Thus, the dichotomy for $\wedge$ applies to $\vee$ as well, only that we characterise the classes with the wildcard matrix in which each element has been negated. This is called the negated wildcard matrix (see Figure~\ref{fig:bad-matrix}).

We now show that the equality operator ``$=$'' requires $\Omega(m)$ bits of space. First note that the output from the pattern matching algorithm is 0 if and only if $\Delta([P[j],T[i+j])=0$ for an odd number of positions $j$. For example, if $M_\Delta$ is the identity matrix, $\local_{(=,\Delta)}$ gives us the parity of the Hamming distance.

Since $\local_{(=,\Delta)}$ is valid, there are $x\in\SigmaP$, $a,b\in\SigmaT$ such that $\Delta(x,a)=0$ and $\Delta(x,b)=1$. We reduce from the \indexing problem, where Alice has a string in $\{a,b\}^m$ and Bob has an index $n$. Alice initialises a pattern matching algorithm on the pattern $P=\{x\}^m$ and feeds it with her string. She sends the internal state to Bob, who feeds the algorithm with $n$ copies of the symbol $a$. The first position of $P$ is now aligned with the $n$th character of Alice's string. Suppose the output from the algorithm is $d$. Bob now feeds in another $a$. Let $d'$ be the new output. If $d=d'$ then the character at position $n$ of Alice's string must have been $a$. If $d\neq d'$ then the character must have been $b$.

The operator ``$\neq$'' is similar to ``$=$'' and also requires $\Omega(m)$ bits of space. To see this, note that the output from the pattern matching algorithm is 0 if and only if $\Delta([P[j],T[i+j])=1$ for an even number of positions $j$. We may therefore prove the lower bound using a reduction from the \indexing problem similar to above.

\section{The $L_{\infty}$ distance}\label{sec:infty}

In this section we consider the $L_{\infty}$ distance problem which can be defined as $\local_{(\max,\Delta)}$, where $\Delta(x,y)=|x-y|$ and $\max(a,b)$ is the maximum of $a$ and $b$. In this section we assume that the pattern and text are integer valued. Here the distance function is the maximum $\Delta(P[j],T[i+j])$ over all $j$, that is
\begin{equation*}
    d(i)= \max_{j\in\{0,\dots,m-1\}} \Delta(P[j],T[i+j])\,.
\end{equation*}

\begin{theorem}
    \label{thm:maximum}
    The $L_{\infty}$ distance problem requires $\Omega(m)$ bits of space.
\end{theorem}
\begin{proof}
    Let $\SigmaP=\{0,1\}$ and $\SigmaT=\{2,3\}$. Therefore $\Delta$ is specified according to Figure~\ref{fig:infinity}. Let $\Delta'(x,y)=1$ if $\Delta(x,y)<3$,
    otherwise $\Delta'(x,y)=0$. Therefore $M_{\Delta'}$ contains the wildcard
    matrix and hence by Theorem~\ref{thm:and-linear}, $\local_{(\wedge,\Delta')}$ requires $\Omega(m)$ space.

    Let $d'(i)$ be the distance under $\local_{(\wedge,\Delta')}$. If $d'(i)=1$ then for all $j$, $\Delta'(P[j],T[i+j])=1$, implying that $\Delta(P[j],T[i+j])<3$ for all $j$. Hence $d(i)<3$. If $d'(i)=0$ then there exists a $j$ such that $\Delta'(P[j],T[i+j])=0$, implying that $\Delta(P[j],T[i+j])=3$ and hence $d(i)=3$. Therefore, if we can solve $\local_{(\max,\Delta)}$, we can solve $\local_{(\wedge,\Delta')}$.
\end{proof}

\begin{figure}[t]
    \begin{equation*}
        \begin{array}{c|cc}
            \Delta & 2 & 3\\
            \hline 1 & 1 & 2\\
            0 & 2 & 3
        \end{array}
        \qquad \qquad
        \begin{array}{c|cc}
            \Delta' & 2 & 3\\
            \hline 1 & 1 & 1\\
            0 & 1 & 0
        \end{array}
    \end{equation*}
    \caption{\label{fig:infinity}$\Delta$ and $\Delta'$ in the proof of Theorem~\ref{thm:maximum}.}
\end{figure}

\section{Non-binary alphabets}\label{sec:nonbinary}

The space lower bounds we have given so far have been either $\Omega(\log{m})$ or $\Omega(m)$ bits. When the pattern or text alphabet is drawn from a large universe, the question arises as to whether even more space is required to perform online pattern matching.  We show by way of another different reduction a method that may be applicable to a wider range of pattern matching problems than we consider here. Our approach is to show a reduction from  the communication complexity problem \disjointness~\cite{Kushilevitz:97} to the Hamming distance problem.  In \disjointness Alice and Bob both have sets of $m$ elements each chosen from a universe of size $U$ and Bob wants to determine if their intersection is empty. The lower bound for the space complexity of the Hamming distance problem will then be determined by lower bounds for the one-way randomised communication complexity of the \disjointness problem with private coins.  A result regarded as folklore shows that this complexity is $\Omega(m\log{m}+\log{\log{U}})$ when $U$ is  $\Omega(m^{1+\epsilon})$~\cite{Nisan:2011,P-folklore:09}. This in turn implies a superlinear lower bound for the space complexity of the online Hamming distance problem with large alphabets.

For an integer $n$, we write $[n]$ to denote the set $\{0,\dots,n-1\}$. Alice has a set $A\subseteq [U]$ and Bob has a set $B\subseteq [U]$, and $|A|=|B|=m$. The reduction performs the following steps. We assume for the moment that Alice and Bob both have a shared source of randomness and show later how this assumption can be removed.

\begin{enumerate}
 \item Alice creates a pairwise independent hash function $h:[U]\rightarrow [cm]$, for some constant integer $c>1$ and creates a pattern $P$ of length $cm$ where each element is initialised to be some unique symbol $\$\notin [U]$. She then sets $P[h(x)] = x$ for all $x\in A$ by going through $A$ in some arbitrary order. If a position of $P$ is written to multiple times, only the last write is stored.
 \item Alice starts the Hamming distance algorithm up until the point at which it has processed the pattern $P$ but none of the text (which is created later) and sends the internal state of the algorithm to Bob.
 \item Bob performs the same hashing operation using the same hash function but this time on set $B$, creating a text $T$ of length $cm$. Bob uses a different unique symbol $\$'\notin [U]$ for the initialisation of the text.
 \item Bob feeds the Hamming distance algorithm with the whole text $T$. Bob concludes that $A$ and $B$ are disjoint iff the output is $cm$.
\end{enumerate}

\begin{theorem}
    Any randomised algorithm for Hamming distance where the symbols are chosen from a universe of size $\Omega(m^{1+\epsilon})$ uses $\Omega(m\log{m} + \log{\log{U}})$ bits of space.
\end{theorem}
\begin{proof}
    Considering the reduction above, if $A$ and $B$ are disjoint, then a deterministic Hamming distance algorithm will always output $cm$. If $A$ and $B$ are not disjoint then a necessary condition for a deterministic Hamming distance algorithm to output $cm$ is if at least two elements are hashed to the same location by either Alice or Bob. We can see that the probability of incorrectly outputting $cm$ is maximised when $A$ and $B$ share exactly one element. Therefore, suppose that $A\cap B=\{x\}$. The element $x$ is hashed to position $h(x)$. By the union bound and the pairwise independence of the hash function, the probability that some other element in either $A$ or $B$ is mapped to $h(x)$ is at most $1/(cm)\cdot m\cdot 2=2/c$. If we assume our randomised Hamming distance algorithm is correct with probability at least $2/3$, then the overall process falsely reports disjointness with probability at most $2/c + 1/3$ (union bound).  The space complexity of Hamming distance is therefore lower bounded by the communication complexity of the disjointness problem if Alice and Bob have a shared source of random bits to select their common hash function. By Newman's Theorem~\cite{Newman:1991} the cost of transforming the protocol to work with only private coins is at most an additive $O(\log{\log{U}})$ factor in the asymptotic complexity. Assuming that $U$ grows polynomially in $m$ and so $\log{\log{U}}$ is $O(\log{m})$, the overall lower bound for the space complexity of the Hamming distance problem is therefore $\Omega(m\log{m} - \log{m}) = \Omega(m\log{m})$.  To finish the proof for larger $U$, we observe first that a lower bound for smaller universes must still hold for larger ones. The final additive $\Omega(\log \log{U})$ term is derived by simply setting $m=1$ and follows directly from the randomised lower bound for \equality. Therefore the overall lower bound is $\Omega(m\log{m} + \log{\log{U}})$ as required.
\end{proof}

\section{Non-local pattern matching} \label{sec:non-local}

So far we have focused only on local pattern matching where each position in the alignment contributes to the distance independently of the other positions. Here we take a brief look at space lower bounds for two non-local distance measures: edit distance and swap matching.

In online pattern matching, we define the \emph{edit distance}  as the minimum number of single character edit operations (insert, delete and replace) required to transform $P$ into the last $m$ characters of the streamed text. This implies that the number of insertions and deletions are equal.

We show that for binary $\SigmaP=\SigmaT=\{0,1\}$, the online edit distance problem requires $\Omega(m)$ bits of space. For non-binary inputs there is a reduction from the Hamming distance problem~\cite{ApproxEdit:04}. The reduction we give covers the binary alphabet case as well and follows directly from \indexing, where Alice has a string $P\in\{0,1\}^m$ and Bob has an index $n$. Alice initialises a pattern matching algorithm on the pattern $P$ and sends the internal state to Bob, who first feeds in $m$ zeros. Let $d$ be the output and note that $d$ is the number of ones in $P$. Bob then feeds in the $m$-length string that consists of zeros at every position except for at position $n$ where it is one. Let $d'$ be the output. Bob can now decide the value of $P[n]$ by comparing $d$ with $d'$: $P[n]=1$ if $d'<d$, and $P[n]=0$ if $d'\geq d$. The probability of error is therefore upper bounded by the union bound on $d$ and $d'$ being wrong.

Given a string $S$, a \emph{swap} at position $i$ means that the characters $S[i]$ and $S[i+1]$ swap positions. We say there is a \emph{swap match} if and only if the pattern $P$ can be transformed into the last $m$ characters of the streamed text through a set of swaps. Each $S[i]$ is swapped at most once.

We show that the online swap distance problem requires $\Omega(m)$ space. Our proof is based on the techniques we have presented in this paper.  Specifically, we demonstrate a reduction from $\local_{(\wedge,\Delta)}$ where $M_\Delta$ contains the wildcard matrix, hence the space lower bound is $\Omega(m)$. Suppose we have $\Delta$ as in Figure~\ref{fig:swap}. Let $P\in\{\wildcard,x\}^m$ and $\SigmaT=\{a,b\}$. From $P$ we obtain $P'\in\{0,1\}^{5m}$ such that every $\wildcard$ in $P$ is replaced with $00100$ and every $x$ is replaced with $00010$. When we receive characters from the text, we replace $a$ with $00010$ and $b$ with $01000$. It follows, under the transformation of the symbols, that there is a swap match if and only if $\local_{(\wedge,\Delta)}$ outputs \TRUE for the original (non-transformed) strings. To see this, note that both $a$ and $b$, under the transformation, swap match $\wildcard$, but $b$ does not swap match $x$ (see Figure~\ref{fig:swap}). The transformation of the symbols does not allow swaps between adjacent characters; every possible swap will take place ``within'' the binary encoding of a symbol. Thus, a swap match directly corresponds to a match under $\local_{(\wedge,\Delta)}$.
\begin{figure}[t]
    \begin{equation*}
        \begin{array}{c|cc}
             \Delta & a & b\\
             \hline \wildcard & 1 & 1\\
             x & 1 & 0
        \end{array}
        \qquad\qquad\quad
        \begin{array}{cc}
            a: & 00{\bf 01}0\\
            \wildcard: & 00{\bf 10}0
        \end{array}
        \qquad
        \begin{array}{cc}
            b: & 0{\bf 10}00\\
            \wildcard: & 0{\bf 01}00
        \end{array}
        \qquad
        \begin{array}{cc}
            b: & 01000\\
            x: & 00010
        \end{array}
    \end{equation*}
    \caption{\label{fig:swap}$\Delta$ and alignments under swaps.}
\end{figure}

\section{Open problems}

We have considered space lower bounds and discussed how they can be derived from known communication complexity lower bounds.  Upper bounds can also be directly derived from existing online pattern matching algorithms.  For all the problems we have discussed there is at most a log factor gap between these upper and lower bounds.  However, where the known lower bound is sublinear, as is the case for exact matching for example, this gap may still be considered significant. Further, for bounded Hamming distance where the distance is only to be given if it is at most some constant $k$, the best known randomised online space upper bound is $O(k^3 \text{polylog }{m})$~\cite{Porat:09}).  The best known lower bound, on the other hand, is very different at $\Omega(k)$~\cite{HSZZ:06}.  Further, it is known that the lower bounds can not be increased to match the known upper bounds using the one-way communication complexity of the functions between two strings of the same length. Either more space efficient algorithms exist for these problems or novel techniques will be needed to improve the lower bounds.

\bibliographystyle{custom}
\bibliography{bib-latest}

\end{document}